# 7  −  The FHI-aims Code: All-electron, *ab initio* materials simulations towards the exascale

Volker Blum[1], Mariana Rossi[2,3], Sebastian Kokott[4], and Matthias Scheffler[2]

[1] Duke University, Durham, NC, USA
[2] The NOMAD Laboratory at the FHI of the Max-Planck-Gesellschaft and IRIS Adlershof of the Humboldt Universität zu Berlin, Germany
[3] Max-Planck-Institute for the Structure and Dynamics of Matter, Hamburg, Germany
[4] Materials Simulations from First Principles e.V. (MS1P), Berlin, Germany

**Background and Current Status**

FHI-aims is a quantum mechanics software package based on numeric atom-centered orbitals (NAOs) with broad capabilities for all-electron electronic-structure calculations and *ab initio* molecular dynamics. It also connects to workflows for multi-scale and artificial intelligence modeling.

Since its foundation in 2004, the FHI-aims code has been designed with a clear set of goals. It should be numerically precise across the periodic table. It should be "all-electron" (not pseudopotential) and handle periodic systems (i.e., extended models of solids, surfaces, and nanostructures) as well as non-periodic systems (i.e., molecules and clusters). The code should support density-functional theory (DFT) with all relevant exchange-correlation functionals, and it should be amenable to correlated methods beyond DFT, i.e. the random-phase approximation and many-body perturbation theory (e.g., *GW*) based on Green's functions and the screened Coulomb interaction, as well as wave-function based correlation methods from quantum chemistry (MP2 and coupled-cluster theories). Furthermore, the code should scale efficiently from small to very large simulation sizes (thousands of atoms or more) and work seamlessly from limited hardware (laptops) up to the most powerful supercomputers available now or in the future.

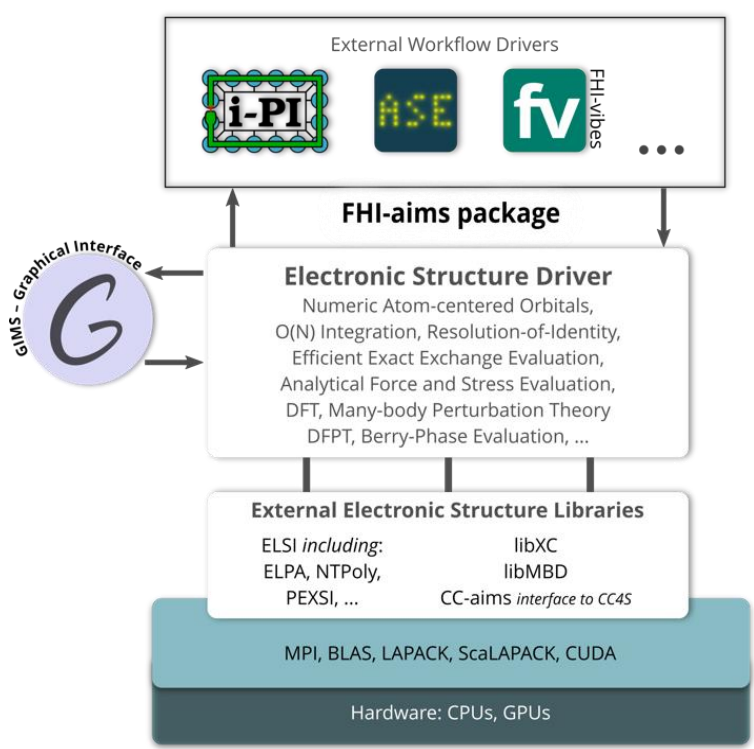

Figure 1: Overview of the FHI-aims code, including its core functionalities, external libraries directly coupled to the code, its integration with external workflow drivers and its integration with the graphical interface GIMS (see Fig. 2).

From the beginnings in 2004, the team grew to include several further key contributors by the time it was first released in 2009 [1]. Today, FHI-aims is a worldwide community project created by well over 150 individual contributors (https://FHI-aims.org/who-we-are) including support for key open source developments such as the ELPA library [2,3] for massively parallel eigenvalue solutions, the ELSI infrastructure for lower-scaling solutions [4], CECAM's electronic structure library [5], environments such as i-PI [6], FHI-vibes [7], the atomic simulation environment ASE [8], the open-source graphical interface for materials science (GIMS) [9], and many others (see Figure 1 and https://fhi-aims.org). The FHI-aims coordinators regularly organize schools and virtual tutorials (available at https://fhi-aims.org). Outreach efforts include industry, through the non-profit association MS1P (https://ms1p.org), ensuring that associated income is returned to the community via code advancements.

**Development Priorities**

The numerical foundation on NAO basis sets lies at the core of FHI-aims, allowing to represent the electronic structure of any problem in chemistry or materials science and engineering, without shape approximations. Support for and compatibility with Gaussian-type and Slater-type orbitals is contained in the code and important for excited-state calculations and electron-electron correlation beyond DFT. Key priorities that drive the ongoing developments include:

- *Usability*. Like many of its peer codes, FHI-aims is usable as a single binary at the command line of a terminal, through queueing systems at supercomputer facilities, or embedded in an ecosystem of separately developed and/or customized scripted tools for higher-level tasks [6,7,8]. The input to FHI-aims itself is simple, requiring only two input files and a few minimal keyword additions to get started; several tutorials and a browser based graphical interface, "GIMS", [9] (Figure 2) are also available. A key ongoing challenge lies in the ever-evolving complexity of high-performance computer systems, especially for the demanding applications of current and urgent interest. In this context, we note that critical tools in high-performance computing, such as message-passing interface (MPI) libraries, compilers and numerical libraries are insufficiently standardized and can present a steep learning curve for newcomers to the field. Reducing this learning curve through tutorials, testing, dedicated code advances and infrastructure will remain an overarching priority for future FHI-aims developments, in particular targeting new accelerator models towards the exascale era (see below).

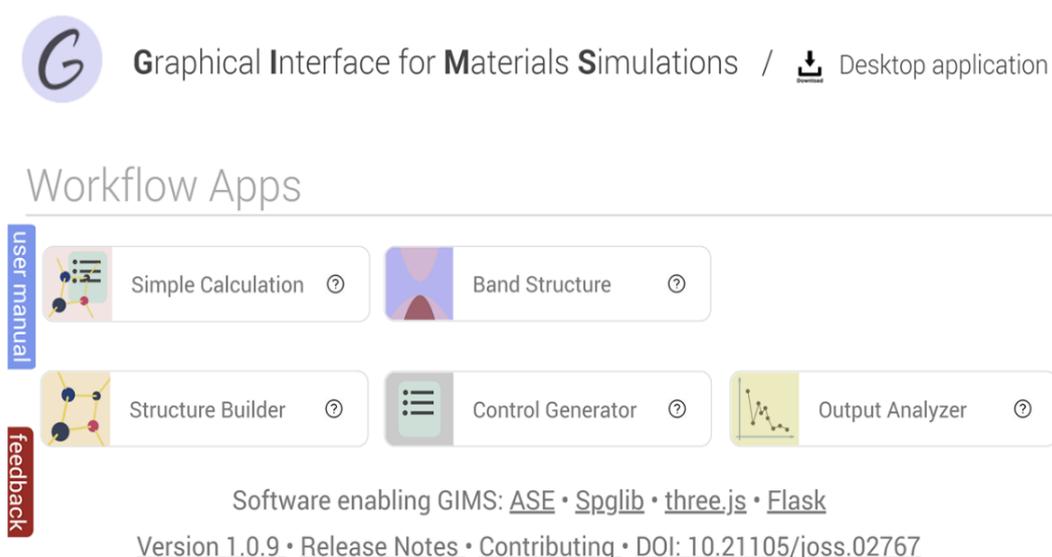

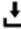

Figure 2: Start page of the Graphical Interface for Materials Simulations (GIMS) [9] for the FHI-aims code (print version). GIMS is completely browser based, i.e., immediately usable on any computer. The interface also supports the *exciting* code and, being built on the ASE [8], it is open to accommodate any other electronic structure code.

- *Community*. FHI-aims is a code based in a large academic community, especially when it comes to a plethora of new developments that a single, small team could not shoulder. Key examples are, e.g., refined density functionals that capture dispersion interactions accurately, real-time time dependent DFT, incorporation of nuclear quantum effects both in the code and by external tools, thermal and electrical transport calculations, *GW* approaches, and many more. Therefore, it is a matter of course to keep the FHI-aims code open, accessible and welcoming to a large community of existing and new users and developers.

- *Science*. In order to keep pace with the increasing needs of our field, continuous work on new features is essential. Examples of our ongoing work include efficient hybrid DFT for 10,000 atoms and more, relativistic formalisms capturing the full Dirac equation, important to capture spin-related phenomena, e.g., in "quantum materials", coupled-cluster theory for high accuracy of stability, reactions in and reactions between extended solids, and a plethora of approaches geared at accurately simulating excitations of the electronic and nuclear systems of a solids that connect to powerful spectroscopies as well as to device applications (e.g., optoelectronic or spintronic) by our experimental colleagues. Connecting the electronic structure foundation to artificial intelligence approaches in order to accelerate computational steps that do not need to repeated and/or can be predicted based on already existing information is a critical practical step for all these objectives.[10]

**Meeting the Exascale Challenges**

Many of our ongoing developments aim at enabling investigations of systems of higher complexity, systematic consideration of metastable states and temperature, and all this at significantly (urgently needed) higher accuracy than what is possible today. Importantly, the goal of utilizing ever-faster computing architectures goes beyond 'speeding up' state-of-the-art high-throughput computations that still employ the theory of the 1990s (through widely used, successful, but also fundamentally limited semilocal density functionals).  Figure 3 shows the schematic reach of different levels of electronic structure theory; in FHI-aims, high-accuracy approaches to electronic structure theory are expected to benefit most directly from the exascale hardware.

Exascale architectures will be heterogeneous, featuring both CPUs as well as accelerators such as GPUs - the latter coming in various flavors (at the time of writing, at least NVidia, AMD, and Intel) and with different coding paradigms. Our strategy in FHI-aims has been to build the code around an "MPI first" paradigm, meaning that every computational method is foremost parallelized without any *a priori* restriction of execution across compute nodes in even the largest supercomputers (~200,000 cores were demonstrated already in 2011). Shared-memory parallelism within each node can be implemented through the MPI-3 standard for multicore processing systems where needed, but importantly controllable where needed from within the code. Through work for NVidia GPUs, we already have a working strategy to treat particular computational hotspots by GPU offloading [2,3,11]. Figure 4 shows the impressive power that can be leveraged for large problems on even a few nodes of the pre-exascale computer Summit (42 Power9 cores and six NVidia V100 GPUs per node) at Oak Ridge National Laboratory, compared to many tens of nodes of the Cori Intel Haswell system (32 cores per node) at National Energy Research Center only a few years earlier. Ongoing work focuses on extending this paradigm throughout the code as well as to the newer AMD and Intel architectures.

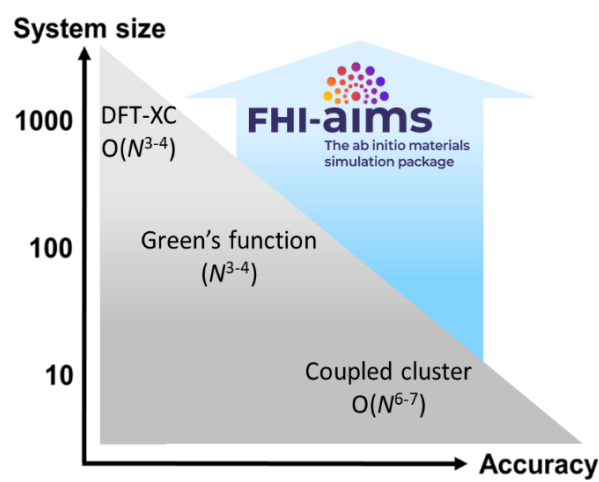

Figure 3: Schematic scaling of various electronic-structure methods and system size (# of atoms) currently possible. For DFT, we indicate cubic scaling since lower scaling is typically not yet reached for ~1,000 atoms in dense systems, in our experience. Direct electronic-structure calculations for mid- to large-scale systems at the highest levels will benefit most dramatically from successful exascale implementation. Our efforts in FHI-aims will focus especially on these highest levels, e.g., in work done in the NOMAD Center of Excellence (https://www.nomad-coe.eu).

FHI-aims already includes advanced exchange-correlation methodologies such as the random-phase approximation, second-order Møller-Plesset perturbation, and coupled-cluster theory. These are presently being extended for the exascale hardware. For instance, the scalability and performance of large-scale DFT calculations is determined by the eigensolver. For hybrid DFT calculations, a second bottleneck is the evaluation of the non-local exact-exchange part of the Fock matrix, and for *GW*, RPA, and CC calculations it is determined by algebraic tensor operations. We are tackling these challenges together with wider community efforts, e.g., ELPA [2,3], ELSI [4], and the NOMAD Center of Excellence (https://www.nomad-coe.eu).

For large systems, the time spent for diagonalization in DFT is always a potential bottleneck. $O(N^3)$ ($N$: system size) scaling dense linear algebra approaches remain competitive with alternatives up to thousands of atoms in our benchmarks. We are therefore helping to enhance the eigensolver ELPA in terms of functionality, performance and energy efficiency, in order to deliver an exascale version. Through the ELSI infrastructure project,[5] we are also connected to other highly efficient solvers that scale lower than $O(N^3)$, such as NTPoly, $O(N)$, or the PEXSI solver, $O(N^2)$. Google's Tensor Processing Units (TPUs) were recently employed to accelerate FHI-aims' conventional DFT (no sparsity assumptions) to almost 250,000 orbitals [12].

When non-local operators are needed (e.g., for hybrid functionals), the bottleneck is created by the formally (without accounting for sparsity) quartic scaling of the method. For hybrid DFT, O($N$) scaling has long been realized, but overhead remains especially for intermediate-sized systems. For the even more challenging beyond-DFT methods, we are working on providing low-scaling, efficiently load-balanced implementations for RPA, MP2, and *GW*, using the real-space and imaginary time treatment or variations thereof. This will include, most critically, sparse matrix-matrix operations, batched matrix-matrix multiplications, and data rearrangement.

**Concluding Remarks**

FHI-aims is used and advanced by a great community. The code has been used for a wide range of calculations and, via workflows, many multi-scale modeling and artificial intelligence analyses (see e.g. Ref. 10). Pre-exascale architectures are already well supported. In addition to "heroic" largest-scale calculations, FHI-aims is also capable of launching an essentially unlimited number of separate, ensemble-parallel calculations at once via split MPI communicators, a mode of operation that is well suited for the exascale regime. Exascale computing may have a significant energy footprint. Here the FHI-aims community works on systematic optimization, e.g. by active learning strategies and workflows that start from the knowledge of the NOMAD data base (https://nomad-lab.eu/services/repo-arch) and make educated decisions for special DFT calculations in order to create a reliable and informative data pool for a faithful AI description. Statistical mechanics and multi-scale modeling require long time and length scales and here, for example, the hand-shake linkage to machine-learned potentials is being developed.

**Acknowledgements**

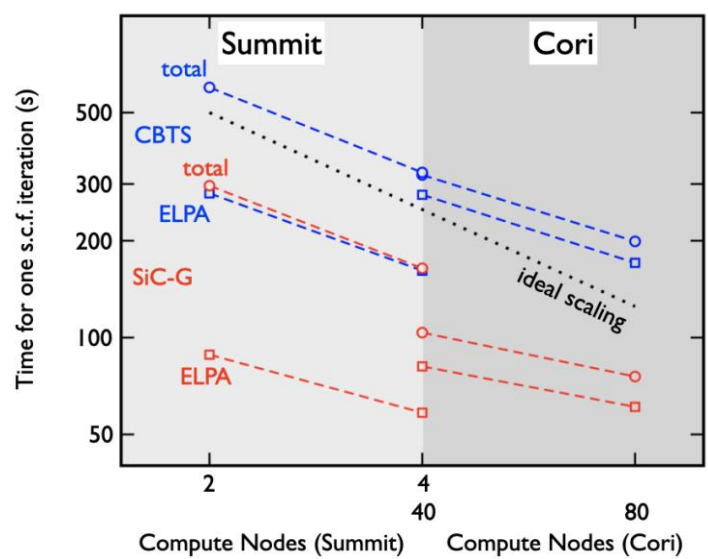

Figure 4: Visualization of computational time as a function of number of compute nodes used, for one self-consistent field (s.c.f.) iteration of semilocal DFT for two large systems on the Cori supercomputer (CPU only, 40 and 80 nodes, right side) and on the Summit supercomputer (CPU+GPU, 2 and 4 nodes, left side), showing previously published data from Table 1 of Ref. [3]. FHI-aims' "light" settings were used. "CBTS" (blue curves) is a 3,000-atom periodic supercell model of a $Cu_2BaSnS_4$ semiconductor. "SiC-G" is a 3,376-atom slab model of a graphene layer on a SiC(111) substrate. Data is shown for the total time per s.c.f. iteration (circles) and for the portion consumed by the eigenvalue solver ELPA. On Summit, a portion of the s.c.f. cycle not related to the eigensolver (the electrostatic potential) is not yet accelerated on GPUs.

Foremost we thank the wide community of FHI-aims users and developers, the latter are listed at https://fhi-aims.org/who-we-are. We do not have space to list everyone individually, but we would like to particularly thank the Max Planck Computing and Data Facility (Garching, Germany),  for their continued support. We thank Bernard Delley for several discussions that started well before the FHI-aims developments began and continued afterwards. We also thank Martin Fuchs for key discussions and software contributions during the initial development phase and Jörg Behler and Karsten Reuter for advice during the inception of FHI-aims. This work received funding from the European Union's Horizon 2020 Research and Innovation Programme (grant agreement No. 951786, the NOMAD CoE), and the ERC Advanced Grant TEC1P (No. 740233). Work on ELSI, ELPA-GPU and the CECAM ESL was partially supported by the National Science Foundation under Award Number 1450280.

## References

[1] V. Blum, R. Gehrke, F. Hanke, P. Havu, V. Havu, X. Ren, K. Reuter, M. Scheffler, Ab initio molecular simulations with numeric atom-centered orbitals. Comput. Phys. Commun.**180** (2009) 2175-2196. https://doi.org/10.1016/j.cpc.2009.06.022.

[2] A. Marek, V. Blum, R. Johanni, V. Havu, B. Lang, T. Auckenthaler, A. Heinecke, H.-J. Bungartz, H. Lederer, The ELPA library: scalable parallel eigenvalue solutions for electronic structure theory and computational science. J. Phys.: Condens. Matter **26** (2014) 213201. https://doi.org/10.1088/0953-8984/26/21/213201.

[3] V. W.-z. Yu, J. Moussa, P. Kus, A. Marek, P. Messmer, M. Yoon, H. Lederer, V. Blum, GPU-Acceleration of the ELPA2 Distributed Eigensolver for Dense Symmetric and Hermitian Eigenproblems. Comput. Phys. Commun. **262** (2021) 107808. https://doi.org/10.1016/j.cpc.2020.107808.

[4] V. W.-z. Yu, C. Campos, W. Dawson, A. García, V. Havu, B. Hourahine, W. P. Huhn, M. Jacquelin, W. Jia, M. Keçeli, R. Laasner, Y. Li, L. Lin, J. Lu, J. Moussa, J. E. Roman, A. Vazquez-Mayagoitia, C. Yang, V. Blum, ELSI – An Open Infrastructure for Electronic Structure Solvers. Comp. Phys. Commun. **256** (2020) 107459. https://doi.org/10.1016/j.cpc.2020.107459.

[5] M. J. T. Oliveira, N. Papior, Y. Pouillon, V. Blum, E. Artacho, D. Caliste, F. Corsetti, S. de Gironcoli, A. M. Elena, A. García, V. García-Suárez, L. Genovese, W. P. Huhn, G. Huhs, S. Kokott, E. Küçükbenli, A. H. Larsen, A. Lazzaro, I. Lebedeva, Y. Li, D. López-Durán, P. López-Tarifa, M. Lüders, M. A. L. Marques, J. Minar, S. Mohr, A. A. Mostofi, A. O'Cais, M. C. Payne, T. Ruh, D. G. A. Smith, J M. Soler, D. A. Strubbe, N. Tancogne-Dejean, D. Tildesley, M. Torrent, V. W.-z. Yu, The CECAM Electronic Structure Library and the modular software development paradigm. J. Chem. Phys. **153** (2020) 024117. https://doi.org/10.1063/5.0012901.

[6] V. Kapil, M. Rossi, O. Marsalek, R. Petraglia, Y. Litman, T. Spura, B. Cheng, A. Cuzzocrea, R. H. Meißner, D. M. Wilkins, B. A. Helfrecht, P. Juda, S. P. Bienvenue, W. Fang, J. Kessler, I. Poltavsky, S. Vandenbrande, J. Wieme, C. Corminboeuf, T. D. Kühne, D. E. Manolopoulos, T. E. Markland, J. O. Richardson, A. Tkatchenko, G. A. Tribello, V. Van Speybroeck, M. Ceriotti, i-PI 2.0: A universal force engine for advanced molecular simulations. Comput. Phys. Commun. **236** (2019) 214-223. https://doi.org/10.1016/j.cpc.2018.09.020.

[7] F. Knoop T. A. R. Purcell, M. Scheffler, C. Carbogno, FHI-vibes: Ab Initio Vibrational Simulations. J. Open Source Softw. **5**, 2671 (2020). https://doi.org/10.21105/joss.02601.

[8] A. H. Larsen, J. J. Mortensen, J. Blomqvist, I. E. Castelli, R. Christensen, M. Dułak, J. Friis, M. N. Groves, B. Hammer, C. Hargus, E. D. Hermes, P. C. Jennings, P. B. Jensen, J. Kermode, J. R. Kitchin, E. L.

Kolsbjerg, J. Kubal, K. Kaasbjerg, S. Lysgaard, J. B. Maronsson, T. Maxson, T. Olsen, L. Pastewka, A. Peterson, C. Rostgaard, J. Schiøtz, O. Schütt, M. Strange, K. S. Thygesen, T. Vegge, L. Vilhelmsen, M. Walter, Z. Zeng, K. W. Jacobsen, The atomic simulation environment—a Python library for working with atoms. J. Phys.: Condens. Matter **29** (2017) 273002. https://doi.org/10.1088/1361-648X/aa680e.

[9] S. Kokott, I. Hurtado, C. Vorwerk, C. Draxl, V. Blum, M. Scheffler, GIMS: Graphical Interface for Materials Simulations. Journal of Open Source Software **6**, (2021) 2767. https://doi.org/10.21105/joss.02767.

[10] The NOMAD Artificial Intelligence Toolkit: https://nomad-lab.eu/aitoolkit.

[11] W. P. Huhn, B. Lange, V. W.-z. Yu, M. Yoon, V. Blum, GPU-Accelerated Large-Scale Electronic Structure Theory with a First-Principles All-Electron Code. Comput. Phys. Commun. **254** (2020) 107314. https://doi.org/10.1016/j.cpc.2020.107314.

[12] R. Pederson, J. Kozlowski, R. Song, J. Beall, M. Ganahl, M. Hauru, A. G. M. Lewis, S. B. Mallick, V. Blum, G. Vidal, Tensor Processing Units as Quantum Chemistry Supercomputers. https://arxiv.org/abs/2202.01255v2.